\documentclass[aps,prl,twocolumn,showpacs,showkeys,a4]{revtex4-1}

\usepackage{graphicx}
\usepackage{amsmath}
\usepackage{amssymb}
\usepackage{mathtools}

\usepackage[english]{babel}
\usepackage{dsfont}
\usepackage{bbm}
\usepackage{hyperref} 
\usepackage{amsmath, amsthm, amssymb, mathrsfs}
\usepackage{times}

\usepackage[usenames,dvipsnames]{color}

\newtheorem{corollary}{Corollary}
\newtheorem{theorem}{Theorem}
\newtheorem*{algorithm}{Algorithm}

\DeclareMathOperator{\Tr}{\mathrm{Tr}}
\DeclareMathOperator{\tracedistance}{\mathcal{D}}
\DeclareMathOperator{\e}{\mathrm{e}}
\DeclareMathOperator{\iu}{\mathrm{i}}
\DeclareMathOperator{\hiH}{\mathcal{H}}
\newcommand{\haH}
{H}
\DeclareMathOperator{\rank}{\operatorname{rank}}
\DeclareMathOperator{\gaps}{\operatorname{gaps}}
\def\be{\begin{equation}}
\def\ee{\end{equation}}

\newcommand{\bra}[1]{\langle #1|}
\newcommand{\ket}[1]{|#1\rangle}

\newcommand{\ketbra}[2]{| #1 \rangle \langle #2 |}

\newcommand{\proj}[1]{\vert #1\rangle\!\langle#1 \vert}

\newcommand{\normone}[1]{\| #1 \|_1}
\newcommand{\norminf}[1]{\| #1 \|_\infty}

\newcommand{\dd}{\textrm{d}} 
\newcommand{\rr}{\mathbb{R}} 

\newcommand{\rhogibbs}{\rho_{\mathrm{Gibbs}}^S}

\newcommand{\Appendixcountingrigorous}{{A}}
\newcommand{\Appendixdensityinfreemodel}{{B}}
\newcommand{\AppendixRectangularStates}{{C}}
\newcommand{\AppendixSystemvsBath}{{D}}
\newcommand{\AppendixTemperatureInTheAlgorithm}{{E}}
\newcommand{\Appendixalgorithconvergence}{{F}}
\newcommand{\Appendixboixo}{{G}}
\newcommand{\Appendixtheoremonefornonconstantdensityofstates}{{H}}

\begin{document}
\title{Thermalization in nature and on a quantum computer}
\author{Arnau Riera, Christian Gogolin, and Jens Eisert}

\affiliation{Dahlem Center for Complex Quantum Systems, Freie Universit{\"a}t Berlin, 14195 Berlin, Germany}
\affiliation{Institute for Physics and Astronomy, University of Potsdam, 14476 Potsdam, Germany}

\begin{abstract}
In this work, we show how Gibbs or thermal states appear dynamically in closed quantum many-body systems, building on the program of dynamical typicality.
We introduce a novel perturbation theorem for physically relevant weak system-bath couplings that is applicable even in the thermodynamic limit. 
We identify conditions under which thermalization happens and discuss the underlying physics. 
Based on these results, we also present a 
fully general quantum algorithm for preparing Gibbs states on a quantum computer 
with a certified runtime and error bound.
This complements quantum Metropolis algorithms, which are expected to be efficient but have no known runtime estimates and only work for local Hamiltonians.
\end{abstract}


\maketitle

How do thermal quantum states -- cornerstones of a description in canonical ensembles in quantum statistical physics -- arise from the underlying theory of quantum physics?
This question, a long tradition as it obviously has, is in many ways still surprisingly wide open.
Indeed, much progress was made only recently \cite{Popescu06,equilibration,speed,Goldstein06,reimann10,tasaki98,dong07,absence,Yung10,
AnalyticalQuench,Metropolis}; this is motivated and triggered both by new mathematical \cite{equilibration,speed,reimann10,absence,
Concentration,AnalyticalQuench,Metropolis} 
and numerical \cite{NumericalQuench} techniques becoming available, as well as by new  
experiments with quantum many-body systems in nonequilibrium \cite{Bloch}.

In this work we present a set of precise sufficient conditions for the emergence of Gibbs states from the underlying microscopic theory of quantum mechanics. Our results go beyond previous approaches in that they apply in a physically relevant weak coupling limit and constitute the key insight leading to the invention of a quantum algorithm that prepares Gibbs states with certified precision and runtime.

The three ingredients that enter the standard textbook proof of the canonical ensemble in classical statistical physics are: (i) the \emph{equal a priori probability postulate} (also known as microcanonical ensemble) and an equilibration postulate (such as the second law), (ii) the assumption of \emph{weak coupling}, and (iii) an assumption about the \emph{density of states} of the bath, namely, that it grows faster than exponentially with the energy and that it can be locally well approximated by an exponential \cite{landaulifshitz}.
Here each of these steps is translated to the \emph{pure state quantum statistical mechanics} approach \cite{Popescu06,equilibration,speed,Goldstein06,reimann10,absence,tasaki98,dong07}.
In particular (i) can be replaced by either a typicality argument, or a statement about dynamical relaxation that follows directly from quantum mechanics and (ii) is made precise by proving a novel perturbation theorem that has applications far beyond the scope of the present article.

Our new technical results allow us to design a \emph{quantum algorithm preparing Gibbs states} with explicit error and runtime bounds, invoking a new variant of phase estimation.
Our algorithm complements another algorithm with certified runtime that was proposed in Ref.~\cite{Poulin} and recent developments on quantum Metropolis algorithms \cite{Metropolis}.

\paragraph{Setting and notation.}
We consider a system $S$ weakly coupled to an environment $B$. The Hilbert space reads $\hiH=\hiH_S \otimes \hiH_B$, where $\hiH_S$ and $\hiH_B$ are the Hilbert spaces of the subsystem and the ``bath'' (with finite dimensions $d_S$ and $d_B$).
The evolution of the total system is governed by the Hamiltonian $\haH = \haH_0 + V$, with eigenvalues and eigenvectors $\{E_k\}$ and $\{|E_k\rangle\}$ consisting of an uncoupled Hamiltonian $\haH_0 = \haH_S + \haH_B$, with eigenvalues and eigenvectors $\{E_k^{(0)}\}$ and $\{|E_k^{(0)}\rangle\}$, and a coupling Hamiltonian $V$.
We give conditions under which the reduced state $\psi^S_t=\Tr_B \psi_t$, with $\psi_t = |\psi_t\rangle\langle\psi_t|$, of the subsystem $S$ relaxes for most times to a Gibbs state $\rhogibbs \coloneqq \e^{-\beta \haH_S } / \Tr \e^{-\beta \haH_S }$ with inverse temperature $\beta$ under unitary time evolution $|\psi_t\rangle=\e^{-i \haH t}\ket{\psi_0}$.
By this we mean that for most times their trace distance $\tracedistance(\psi^S_t,\rhogibbs)$, which measures the physical distinguishability \cite{footnote0}, is small.
Note that the decomposition of a given $\haH$ into $\haH_S$, $\haH_B$, and $V$ is not unique.
This freedom can be used to optimize the bounds in our results, and the correct $\haH_S$ naturally results from this optimization.
We assume that the Hamiltonians $\haH$ and $\haH_0$ are non degenerate such that \emph{time averaging} and \emph{dephasing} in the eigenbasis give the same result
\begin{equation*}
  \omega \coloneqq \overline{{\psi_t}} = \lim_{T \rightarrow \infty} \frac{1}{T} \int_0^{T} \psi_t \mathrm{d}t = \sum_k \ketbra{E_k}{E_k} \psi_0 \ketbra{E_k}{E_k} .
\end{equation*}
Whenever an expectation value equilibrates, it does so to the expectation value in $\omega$ \cite{footnote0}.

\paragraph{``Natural thermalization'': Conditions for Gibbs states to appear.} 
In this section we go through points (i)--(iii).
The final conclusion is summarized in Corollary~\ref{theorem:summaryofnaturalthermalization}.
The central point of the argument is a novel perturbation theorem that relates spectral projectors of weakly interacting and noninteracting Hamiltonians in a physically relevant weak coupling limit.
It allows us to connect results on dynamical equilibration and measure concentration with classical counting arguments and thereby prove a set of natural sufficient conditions for thermalization in quantum mechanics.

A stepping stone in the argument will be states that have an energy distribution that is flat in an interval $[E,E+\Delta]$ and vanishes otherwise.
We indicate such states, and their dephased states, by a subscript $\sqcap$ like in $\psi_\sqcap$ or $\omega_\sqcap$ and call them \emph{rectangular states}.
This class of states includes both \emph{mixed states} (in particular the microcanonical state $\omega_\sqcap$) and \emph{pure states} and thus, because of the freedom to choose the phases, usually also initial states that can be locally out of equilibrium.

The \emph{equal a priori probability postulate} (i) can be replaced by a typicality argument using results from Refs.~\cite{Popescu06,equilibration,Goldstein06}.
In Ref.~\cite{Popescu06} powerful concentration of measure techniques are used to show that almost all states from a microcanonical subspace corresponding to a microcanonical energy window $[E,E+\Delta]$ locally look like the reduction of the corresponding microcanonical state; i.e., $\tracedistance(\psi^S,\omega_\sqcap^S)$ is small for all but exponentially few of the states $\psi$ from the subspace, where $\omega_\sqcap$ is the microcanonical state on the subspace.
Alternatively one can use the results concerning the dynamics of states with a high effective dimension of Refs.~\cite{equilibration,reimann10}.
Under one assumption on the spectrum of the Hamiltonian (nondegenerate energy gaps), it is shown that all reduced states on small subsystems of such states tend to the time-averaged equilibrium state $\omega^S$ and stay close to it for most times. In many-body systems,  natural initial states have a high effective dimension, and this is provably true for all but exponentially few states from a microcanonical subspace \cite{equilibration}.

\begin{figure}[b]
  \centering
  \includegraphics[width=0.9\linewidth]{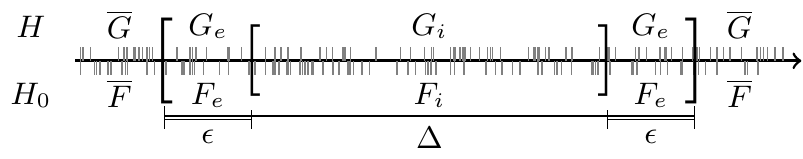}
  \caption{\label{fig:projectors}Definition of the projectors used in the proof of Theorem~\ref{theorem:interactingvsnoninteractingcase}.}
\end{figure}

The delicate issue, which has up to now not been addressed in the literature in a general and rigorous way, is the \emph{weak coupling} approximation (ii) \cite{reimann10,footnote1}.
The problem is that due to the exponential growth of the Hilbert space dimension and the at most polynomial growth of the energy content, the spectrum of the noninteracting Hamiltonian $\haH_0$ becomes exponentially dense with increasing bath size.
Therefore, the \emph{perturbative} limit, in which the coupling $V$ is weak compared to the \emph{gaps} of the noninteracting Hamiltonian $\haH_0$, and in which it can be guaranteed that the energy eigenvectors $\ket{E_k}$ of the full Hamiltonian $\haH = \haH_0 + V$ are close to product states, is arguably \emph{not the physically relevant weak coupling limit}.
Even worse, in this limit memory effects provably prevent thermalization \cite{absence}.
As in the classical setting, a coupling should be considered to be \emph{weak} as long as it does not change the total energy in a \emph{noticeable way}. 
That is to say, the energy stored in the interaction is much less than our (microcanonical) uncertainty about the energy of the system, i.e., $\norminf{V} \ll \Delta$, or for thermalizing systems much less than the thermal energy $1/\beta$.
This is the \emph{relevant weak coupling limit} in which we prove equilibration towards a Gibbs state.
We do this by relating the dephased/microcanonical state $\omega_\sqcap$ to the state $\omega_\sqcap^{(0)}$ dephased with respect to the non interacting Hamiltonian, for which we can easily perform the partial trace to obtain $\omega_\sqcap^{S(0)}$ and thereby an approximation to $\omega_\sqcap^S$.
\begin{theorem}[interacting vs. noninteracting case]
  \label{theorem:interactingvsnoninteractingcase}
  Let $\omega_\sqcap^{(0)}$ and $\omega_\sqcap$ be the dephased/microcanonical states belonging to the interval $[E,E+\Delta]$ with respect to $\haH_0$ and $\haH = \haH_0 + V$; then for every $\varepsilon < \Delta/2$ 
  \begin{align}
    \tracedistance(\omega_\sqcap^{S},\omega_\sqcap^{S(0)}) \leq \tracedistance(\omega_\sqcap,\omega_\sqcap^{(0)}) \leq \frac{\norminf{V}}{\varepsilon} + \frac{\Delta\Omega+\Omega_\varepsilon}{2\,\Omega_{\mathrm{max}}} \, ,
  \end{align}
  where $\Omega_{\mathrm{max}}$ and $\Delta\Omega$ are the maximum, and the difference, of the dimensions of the supports of $\omega_\sqcap^{(0)}$ and $\omega_\sqcap$, and $\Omega_\varepsilon$ is the total number of eigenstates of $\haH$ and $\haH_0$ in the intervals $[E,E+\varepsilon]$ and $[E+\Delta-\varepsilon,E+\Delta]$.
\end{theorem}
The theorem shows that, for any two initial (possibly pure) states that have a flat energy distribution in the interval $[E,E+\Delta]$ with respect to the Hamiltonians $\haH_0$ and $\haH$ with $\norminf{V} \ll \Delta$, the distance of their reduced dephased states $\omega_\sqcap^{S(0)}$ and $\omega_\sqcap^S$ is small.
In particular, assuming an approximately constant density of states such that $\Omega_\varepsilon/(2\,\Omega_{\mathrm{max}}) \approx 2 \varepsilon/\Delta$ and $\Delta\Omega/(2\Omega_{\mathrm{max}}) \lessapprox \norminf{V}/\Delta$, the best choice for $\varepsilon$ is $\varepsilon \approx \sqrt{\norminf{V} \Delta/2}$ which gives
\begin{equation}
\label{eq:intuitive-theorem-1}
  \tracedistance(\omega_\sqcap^{S},\omega_\sqcap^{S(0)}) \lessapprox 4 \sqrt{\frac{\norminf{V}}{\Delta}} .
\end{equation}
In cases with an exponential density of states, for which we will get equilibration towards $\rhogibbs \propto \e^{-\beta \haH_S}$, we can guarantee that $\tracedistance(\omega_\sqcap^{S},\omega_\sqcap^{S(0)})$ is small whenever $\norminf{V} \ll 1/\beta$ (compare Appendix~\Appendixtheoremonefornonconstantdensityofstates{}).
\begin{proof}
  First note that by monotonicity of the trace distance and the triangle inequality
  \begin{equation}
   	\tracedistance(\omega_\sqcap^{S},\omega_\sqcap^{S(0)}) 
	\leq \frac{1}{2}\normone{\omega_\sqcap - \omega_\sqcap^{(0)}} \leq \frac{\normone{G - F} + \Delta\Omega}{2\,\Omega_{\mathrm{max}}} ,
  \end{equation}
  where $G$ and $F$ are the projectors onto the support of $\omega_\sqcap$ and $\omega_\sqcap^{(0)}$ respectively and $\Omega_{\mathrm{min/max}} = \min/\max(\rank(G),\rank(F))$, and $\Delta\Omega = \Omega_{\mathrm{max}} - \Omega_{\mathrm{min}}$.
  It remains to bound $\normone{G - F}$.
  Let $\overline{G} = \mathds{1} - G$ and $\overline{F} = \mathds{1} - F$; then $G - F = G\overline{F} - \overline{G}F$, and thus $\normone{G - F} \leq \normone{G\overline{F}} + \normone{\overline{G}F}$.
  To bound $\normone{G\overline{F}}$ we decompose $G = G_i + G_e$ into an interior part $G_i$ which is the projector onto the eigenstates from the interval $[E+\varepsilon,E+\Delta-\varepsilon]$ and the exterior part $G_e$ and find $\normone{G\overline{F}} \leq \normone{G_i\overline{F}} + \normone{G_e}$ (see Fig.~\ref{fig:projectors}).
  Using the inequality $\normone{\cdot} \leq \rank(\cdot) \norminf{\cdot}$, submultiplicativity of the rank, and that 
 $\rank(G_i) \leq \Omega_{\mathrm{max}}$ this can be recast into $\normone{G\overline{F}} \leq \Omega_{\mathrm{max}} \norminf{G_i\overline{F}} + \rank(G_e)$.
  Finally, from Theorem~V.II.3.1 in Ref.~\cite{bhatia} it follows that $\norminf{G_i\overline{F}} \leq \norminf{V}/\varepsilon$.
  Repeating the argument for $\normone{\overline{G}F}$, introducing the notation $\Omega_\varepsilon = \rank(G_e) + \rank(F_e)$, and putting everything together gives the desired result.
\end{proof}

The level counting argument (iii) -- with is ultimately the reason for the exponential form of $\rhogibbs \propto \e^{-\beta \haH_S}$ -- carries over to the quantum case in a straightforward way in the \emph{absence} of coupling between system and bath \cite{Goldstein06,reimann10}, and with a bit more work one can also obtain a rigorous trace norm error bound.
If the number of states of the bath $\Omega^B_\Delta(E^B)$ in the interval $[E^B,E^B+\Delta]$ is such that the proportion $\Omega^B_\Delta(E-E^S_k)/\sum_l \Omega^B_\Delta(E-E^S_l)$ is close to $\e^{-\beta E^S_k}/\sum_l\e^{-\beta E^S_l}$ for the given $E$ and $\Delta$ and some $\beta$, then the distance of $\tracedistance(\omega_\sqcap^{S(0)},\rhogibbs)$ is small.
This can be guaranteed under a set of natural assumptions that are satisfied by a wide range of natural quantum many-body systems and that resemble the ones commonly used in classical statistical physics, such as an exponential increase of the density of states (Appendix \Appendixcountingrigorous{}).
In particular, for a bath consisting of $m$ noninteracting spin-$1/2$ particles with a slightly varying on site field strength and average local energies of $0$ and $\eta$ one finds (Appendix \Appendixdensityinfreemodel{})
\begin{equation}
  \label{eq:countingargumentbound}
  \tracedistance(\omega_\sqcap^{S(0)},\rhogibbs) \leq \frac{1}{2} (\e^{2 \frac{\norminf{\haH_S}^2}{\eta^2 m}}-1) + C \,
\end{equation}
with $C$ exponentially small in the bath size.
We will later use this bath in our algorithm.
In summary, Eq.~\eqref{eq:countingargumentbound}, Theorem~\ref{theorem:interactingvsnoninteractingcase}, and the results on dynamical equilibration and random states from the unitary invariant measure derived in Refs.~\cite{Popescu06,equilibration} lead to the following conclusions:
\begin{corollary}
  \label{theorem:summaryofnaturalthermalization}
  (Kinematic)
  Almost all pure states $\psi$ from a microcanonical subspace corresponding to an energy interval $[E,E+\Delta]$ of a weakly interacting, sufficiently large quantum system are locally close to a Gibbs state in the sense that for every $\gaps(\haH_0) \ll \varepsilon < \Delta/2$ the probability that
  \begin{equation}
    \tracedistance(\psi^S,\omega_\sqcap^{S(0)}) \geq \frac{2 d_S}{\sqrt{\Omega_{\mathrm{min}}}} + \frac{\norminf{V}}{\varepsilon} + \frac{\Delta\Omega+\Omega_\varepsilon}{2\,\Omega_{\mathrm{max}}} + \varepsilon'
  \end{equation}
  drops of exponentially with $\Omega_{\mathrm{min}}\,\varepsilon'^2$.
  (Dynamic)
  Moreover, if the Hamiltonian in addition has nondegenerate energy gaps \cite{equilibration}, all initial states $\psi_{\sqcap,t=0}$, even those locally out of equilibrium, with a flat energy distribution in the interval, locally equilibrate towards $\rhogibbs$ in the sense that
  \begin{equation}
    \overline{\tracedistance(\psi_{\sqcap,t}^S,\omega_\sqcap^{S(0)})} \leq \frac{d_S}{2\sqrt{\Omega_{\mathrm{min}}}} + \frac{\norminf{V}}{\varepsilon} + \frac{\Delta\Omega+\Omega_\varepsilon}{2\,\Omega_{\mathrm{max}}} .
  \end{equation}
  Both inequalities are robust against deviations from the rectangular distribution. If the bath has an exponentially increasing density of states only a region of bounded variation followed by a sharp cutoff towards higher energies should be sufficient (for details see Appendix~\AppendixRectangularStates{}).
\end{corollary}

\paragraph{``Artificial thermalization'': A quantum algorithm for Gibbs state preparation.}
It follows from Eq.~\eqref{eq:countingargumentbound} and Theorem \ref{theorem:interactingvsnoninteractingcase} that all one has to do to prepare a Gibbs state is to prepare a state close to $\omega_\sqcap$ or $\omega_\sqcap^{(0)}$ on a suitable combination of system plus bath.
This is the central idea behind the quantum circuit shown in Fig.~\ref{fig:qcircuit}, which prepares thermal states without using any knowledge about the eigenstates of the Hamiltonian.

Quantum algorithms that prepare thermal states have several advantages over classical simulation methods:
Quantum Monte-Carlo methods offer a way to, for example, estimate correlation functions of thermal states on a classical computer.
However, such methods are restricted to certain types of Hamiltonians as they suffer from the sign problem.
A procedure that certifiably prepares Gibbs states in a quantum computer does not only overcome the sign problem, but moreover makes it possible to use thermal state in experiments addressing questions of nonequilibrium dynamics in quantum simulators, for example to study quenches.

Our algorithm requires two registers (see Fig.~\ref{fig:qcircuit}).  The first register $R$ consists of $r$ qubits initially in $\ket{0}$ and is used to perform quantum phase estimation. 
The second register $Q$ holds the quantum system plus bath and is put into a rectangular state by performing the following steps:

1. \emph{Initialization}.
The register $Q$ is initialized into the completely mixed state $\rho_1=\frac{1}{d} \sum_{k=1}^{d}\proj{E_k} \otimes \proj{0}^r$.

2. \emph{Partial quantum phase estimation}.  
A new form of quantum phase estimation is performed, which comprises three steps: 
the application of $r$ Hadamard gates on the qubits of $R$, 
the application of $r$ controlled-$U$ operations (with $U$ raised to successive powers of two), and an inverse Fourier transform on $R$.
After this operation, the state of the registers is
\begin{equation}
  \rho_2
  =\frac{1}{d} \sum_{s,s'=0}^{2^r-1} 
  \sum_{k=1}^{d} \alpha_s(\varphi_k)\alpha_{s'}^*(\varphi_k) \ket{E_k}\bra{E_k}\otimes \ket{s}\bra{s'}\, ,
\end{equation}
where $\varphi_k \coloneqq E_k/\norminf{\haH}$
and $\alpha_s(\varphi_k) \coloneqq \frac{1}{2^r} \frac{1-\exp(2\pi\iu(2^r \varphi_k- s))}{1-\exp(2\pi\iu(\varphi_k- s / 2^r))}$. Note that $|\alpha_s(\varphi_k)|^2$ is a probability distribution that becomes more and more peaked around $s/2^r$ as $r$ increases.

3. \emph{Measurement}.
Measuring the first $q$ qubits of $R$, some binary string $s_*$ of length $q$ is obtained and the system is left in the state
\begin{equation}
  \rho_3\propto
  \sum_{s,s'=s_* \Delta_*}^{(s_*+1) \Delta_*} \sum_{k=1}^d \alpha_s(\varphi_k)\alpha_{s'}^*(\varphi_k) 
  \proj{E_k}\otimes \ket{s}\bra{s'}\, ,
\end{equation}
where $\Delta_* \coloneqq 2^{r-q}$ is the number of states of the ancilla register $R$ compatible with the measurement.
By choosing $r$ one can determine the width $\Delta=\norminf{\haH} 2^{-r}\Delta_*$ of the rectangular state that is prepared.
The measured value of $s_*$ determines the energy $E=\norminf{\haH}2^{-q}s_*$ of the rectangular state, and thereby the inverse temperature $\beta$ of the Gibbs state. 
To thermalize the subsystem at some particular temperature, the previous steps must be repeated until the desired energy is measured.
The number of runs increases exponentially with the inverse temperature $\beta$. This prevents us from preparing thermal states at very low temperatures 
(see Appendix \AppendixSystemvsBath).
This is not a deficit of the algorithm, for otherwise {\tt QMA}-hard problems (the quantum analog of NP \cite{localhamiltonianproblem}) could be efficiently solved.
Any general algorithm will presumably have this feature \cite{localhamiltonianproblem}. 
The final state of $Q$ is
\begin{equation}
  \omega_{QC} \coloneqq \Tr_R \rho_3 \propto \sum_{k=1}^{d} \left(\sum_{s=s_* \Delta_*}^{(s_*+1) \Delta_*}|\alpha_{s}(\varphi_k)|^2\right) \proj{E_k}\, .
\end{equation}
For large enough $r$, this state is close to the desired state $\omega_\sqcap$ with $E=\norminf{\haH}2^{-q}s_*$ and $\Delta=\norminf{\haH} 2^{-r}\Delta_*$.
The precise deviation of $\omega_{QC}^S$ from $\rhogibbs$,
\begin{equation}
  \label{eq:twocontributionstotheerror}
  \tracedistance(\omega_{QC}^S,\rhogibbs) \le \tracedistance(\omega_{QC},\omega_{\sqcap}^{(0)}) + \tracedistance(\omega_{\sqcap}^{S(0)},\rhogibbs).
\end{equation}
depends on the density of states of system plus bath.
A good candidate for the bath is the system of $m$ non interacting spin-$1/2$ particles discussed before (Appendix~\Appendixdensityinfreemodel{}) and we give explicit results for the errors and the complexity for this bath:

\begin{figure}[tb]
  \includegraphics[width=\linewidth]{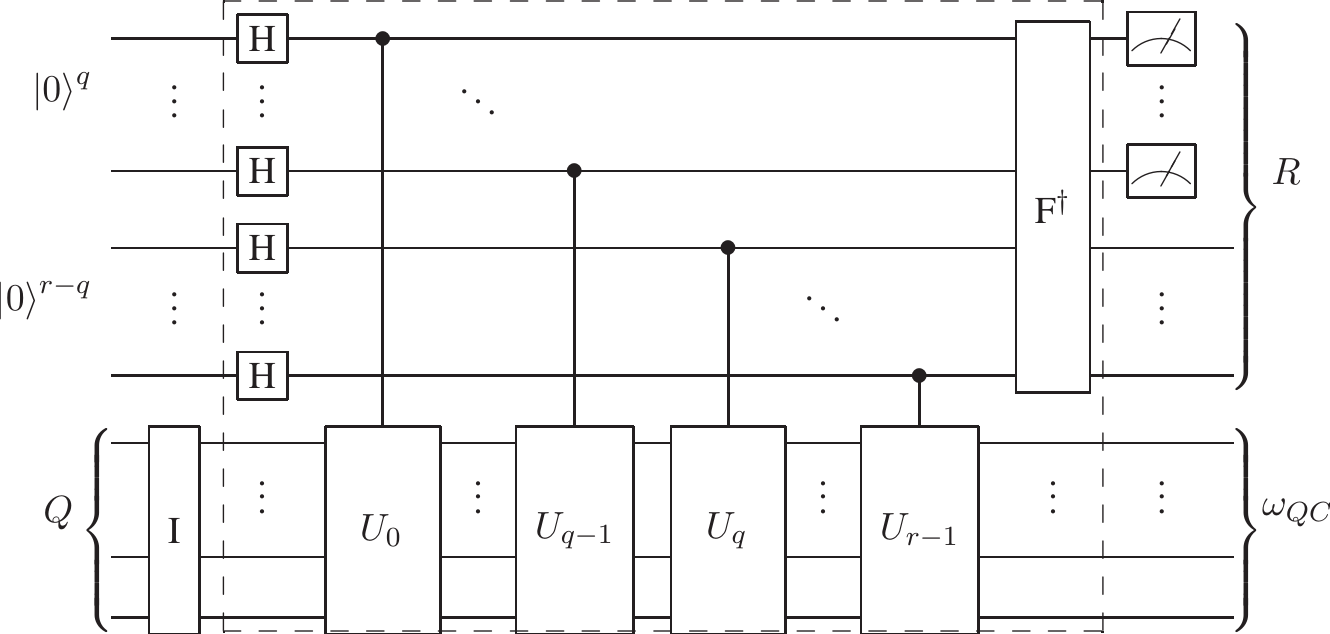}
  \caption{\label{fig:qcircuit}Quantum circuit that generates a dephased rectangular state $\omega^{(0)}_\sqcap$.
   $\mathrm{I}$ is the initialization gate, $\mathrm{H}$ are Hadamard gates, $U_\tau = U^{2^\tau}$, $U = \exp(-\iu \haH_0 /\norminf{\haH_0})$ with $\haH_0 = \haH_S + \haH_B$, and $\mathrm{F}^\dagger$ is the inverse Fourier transform.}
\end{figure}

\begin{algorithm}
  For any chosen $\lambda>0$, any given inverse temperature $\beta$ and system Hamiltonian $\haH_S$, the algorithm presented in Fig.~\ref{fig:qcircuit}, using the bath with $m$ spin-$1/2$ particles and energy scale $\eta=\sqrt{\lambda/m} \norminf{\haH_S}$ discussed before (Appendix \Appendixdensityinfreemodel{}), prepares the system $S$ of $n$ qubits in a state within trace norm distance bounded by
  
  \begin{align}
    \tracedistance(\omega_{QC}^S,\rhogibbs) &\leq 2^{q-r+2} \left(1+\ln(2^{r-q})/\pi^2 \right) \\
    &\times \e^{\frac{2}{\lambda}+\beta \norminf{\haH_S}+\frac{\lambda \norminf{\haH_S}^2 \beta^2}{8} }
      + \frac{1}{2}(\e^{\frac{2}{\lambda}}-1) + C \nonumber
    \end{align}
  with $C$ exponentially small in $m$, 
  to a Gibbs state $\rhogibbs$ with a 
  temperature in the interval $[\beta-\delta\beta,\beta+\delta\beta]$, where
  \begin{equation}
    	\delta\beta \leq 2^{2-q} \sqrt{\frac{\lambda}{m}} 
    	\frac{1}{\norminf{H_S}}\left( 1+\frac{1}{\sqrt{m\lambda}}\right) .
  \end{equation}
  This is achieved using $r$ ancilla qubits and running the algorithm an average number of 
  \begin{equation}
    	\overline{ \sharp \textrm{ runs} } \leq  2^q \sqrt{\frac{\pi }{2 m}} 
    	\e^{\frac{2}{\lambda}+\beta \norminf{\haH_S}+\frac{\lambda \norminf{\haH_S}^2 \beta^2}{8} }
  	\label{eq:number-of-runs}
  \end{equation}
  times, where each run requires the application of $n+2r$ Hadamard gates, $r$ controlled single qubit gates, $n+q$ (with $q\leq r$) single qubit measurements, and $2^r$ controlled unitary time evolutions under $\haH_0 = \haH_S + \haH_B$ for a time $1/\norminf{\haH_0}$.
\end{algorithm}
Notice that the time evolution under $\haH_B$ can be implemented with $m$ gates as the bath is a model of uncoupled spins.
In practice, in absence of an oracle for the Hamiltonian of the system, the error produced to perform the $U$ gate carries a second source of error that comes from the Trotter-Suzuki approximation.
Nevertheless, this error can be suppressed at a polynomial cost for local Hamiltonians \cite{Berry,Poulin}.

The two contributions to the trace distance error \eqref{eq:twocontributionstotheerror} are computed in Appendices~\Appendixdensityinfreemodel{} and \Appendixalgorithconvergence{}, the average number of runs is computed in Appendix~\Appendixalgorithconvergence{} and the error in the temperature comes from the discrete nature of the energy measurement via quantum phase estimation and is calculated in Appendix~\AppendixTemperatureInTheAlgorithm{}.
As is clear form Fig.~\ref{fig:qcircuit}, we need $\sum_{\tau=0}^{r-1} 2^\tau = 2^r$ of the $U$ gates, and the part of the circuit that does not correspond to the controlled time evolution, i.e., the initialization and the inverse Fourier transform,
only requires the implementation of $n+m+2r$ Hadamard gates, 
$r$ controlled single qubit gates,
and $n+m+q$ single qubit measurements.

For a fixed large $\lambda$, the error of the algorithm can be made small by choosing 
an ancilla register of size $O(\beta^2\norminf{\haH_S}^2)$.
The exponential scaling of the runtime with $\beta \norminf{H_S}$ caused by this is not a deficit of the algorithm, as any general efficient algorithm would contradict hardness results such as the local Hamiltonian problem \cite{localhamiltonianproblem}.
Unlike our approach, Metropolis algorithms \cite{Metropolis} are expected to be efficient in some cases, but no rigorous runtime estimates are known and they are only applicable to local Hamiltonians.
An interesting step towards constructing an efficient (in system size) and certified algorithm for local systems was recently made in Ref.~\cite{bilgin} (compare Appendix~\Appendixboixo{}).

\paragraph{Conclusions.}
A set of sufficient conditions for thermalization in quantum mechanics has been presented.
The conditions are a natural translation of the standard assumptions from classical statistical physics.
Along the way, a perturbation argument for realistic weak coupling has been proven that we expect to have significant applications beyond the scope of this Letter.
By using our technical results we are able to design a quantum algorithm preparing thermal states of arbitrary Hamiltonians with rigorous runtime and error bounds.
%
%

\paragraph{Acknowledgments.}
We would like to thank M.\ P.\ M{\"u}ller for discussions, P.\ Reimann for comments, the anonymous referee A for the well justified criticism, 
and the EU (Qessence, Compas, Minos), and the EURYI for support.


\appendix
\section{\Appendixcountingrigorous{} Reduced density matrix of the dephased rectangular decoupled state -- the general case}
In this section the distance between the subsystem of the dephased rectangular decoupled state
and the Gibbs state $\tracedistance(\omega_\sqcap^{S(0)},\rhogibbs)$ is computed.
The discussion is similar to that in Ref.~\cite{reimann10}.
As all eigenvectors of $\haH_0$ are given by tensor products of the eigenstates of $\haH_S$ and $\haH_B$, 
$\ket{E_n}=\ket{E_k^S}\otimes\ket{E_q^B}$,
it is possible to trace the degrees of freedom of the bath of the decoupled rectangular state to get
\begin{equation}
\omega_\sqcap^{S(0)}=\Tr_B \omega_\sqcap^{(0)}
=\sum_{k=1}^{d_S} p_k\proj{E_k^S}\, ,
\end{equation}
with 
\begin{equation}
p_k \coloneqq \frac{\Omega_\Delta^B(E-E_k^S)}{\sum_{k=1}^{d_S} \Omega_\Delta^B(E-E_k^S)}
\, .
\label{eq:exact-probabilities}
\end{equation}
Here, $\Omega_\Delta^B(E-E_k^S)$ denotes the number of states of 
the bath with an energy contained in the interval ${[}E-E_k^S,E-E_k^S+\Delta{]}$.
For simplicity we assume that $\norminf{\haH_S} \le E$.
Notice that the probabilities $\{p_k\}$ only depend on the shape of the spectrum of the bath.
We aim at bounding the trace distance
\begin{equation}
\tracedistance(\omega_\sqcap^{S(0)},\rhogibbs)=\frac{1}{2} \sum_{k=1}^{d_S} |p_k -q_k| \, ,
\end{equation}
where 
\begin{equation}
q_k \coloneqq \frac{\e^{-\beta E_k^S}}{Z_Q}\, ,
\end{equation}
are the eigenvalues of the Gibbs state and $Z_Q \coloneqq \sum_{i=1}^{d_S}\e^{-\beta E_i^S}$.

Let us define a function $S$ as the logarithm of the number of states of the bath
\begin{equation}
S(E) \coloneqq \log \Omega_\Delta^B (E) \, .
\end{equation}
The Hilbert space of the bath is finite dimensional and thus the spectrum of the bath is discrete, from now on it is however assumed that the bath is large enough such that $S:\rr^+\rightarrow \rr^+$ can be well approximated by a twice differentiable function $s:\rr^+\rightarrow \rr^+$.
For natural systems and reasonable energy ranges the additional error from this approximation is exponentially small in the size of the bath.
For the sake of conciseness we ignore such exponentially small errors.
We discuss this continuous approximation more closely in Appendix \Appendixdensityinfreemodel{}.

From now on we will hence focus on $s(E) \coloneqq \log \Xi_\Delta^B(E)$, where $\Xi_\Delta^B$ it a twice differentiable approximation of $\Omega_\Delta^B$.
Taylor's theorem ensures that for every $k$ there exists some $\xi_k \in [E-E_k^S,E]$, such that
\begin{align}
s(E-E_k^S)&= s(E) - \frac{\partial s}{\partial E}(E) E_k^S + \frac{1}{2} \frac{\partial^2 s}{\partial E^2}(\xi_k) {E_k^S}^2 \nonumber \\
&=s(E)-\beta E_k^S +\gamma_k \, ,
\end{align}
where 
\begin{equation}
	\beta \coloneqq \frac{\partial s}{\partial E}(E)
\end{equation}	
is the inverse temperature and
\begin{equation}
	\gamma_k \coloneqq \frac{1}{2} \frac{\partial^2 s}{\partial E^2}(\xi_k) {E_k^S}^2.
\end{equation}	
A linear expansion of $s(E-E_k^S)$ is equivalent to an exponential fit of the smoothed number of states of the bath $\Xi_\Delta^B(E-E_k^S)$,
\begin{equation}
	\Xi_\Delta^B(E-E_k^S)= \e^{s(E)-\beta E_k^S+\gamma_k}\, .
\end{equation}
Thus, the probabilities $\{p_k\}$ can be written as
\begin{equation}
	p_k=\frac{\Xi_\Delta^B(E-E_k^S)}{\sum_{i=1}^{d_S} \Xi_\Delta^B(E-E_i^S)}=
	\frac{\e^{-\beta E_k^S + \gamma_k}}{Z_P}\, ,
\end{equation}
where $Z_P = \sum_{i=1}^{d_S} \e^{-\beta E_i^S + \gamma_i}$.
Therefore, the distance between $\omega_\sqcap^{S(0)}$ and the Gibbs state
depends on how well the density of states of the bath can be approximated by an exponential curve. 
The difference between the probabilities reads
\begin{equation}
	p_k- q_k= \underbrace{\frac{\e^{-\beta E_k^S}}{Z_Q}}_{q_k} \left( \frac{Z_Q}{Z_P} \e^{\gamma_k}-1\right)\, ,
\end{equation}
where the fraction $Z_Q/Z_P$ can be rewritten as
\begin{equation}
	\frac{Z_Q}{Z_P}=\sum_{k=1}^{d_S} \frac{\e^{-\beta E_k^S +\gamma_k}}{Z_P}\e^{-\gamma_k}=\sum_{k=1}^{d_S} p_k \e^{-\gamma_k}\, .
\end{equation}
Introducing the notation 
\begin{align}
  \gamma_{\mathrm{min}} &\coloneqq \min_k \min_{\xi_k\in[E-E_k^S,E]} \gamma_k(\xi_k), \\ 
  \gamma_{\mathrm{max}} &\coloneqq \max_k \max_{\xi_k\in[E-E_k^S,E]} \gamma_k(\xi_k) ,
\end{align}
we can write
\begin{align}
|p_k- q_k| &= q_k\left|  \left( \sum_{k=1}^{d_S} p_k \e^{-\gamma_k} \right) \e^{\gamma_k}-1\right| \nonumber\\
& \le q_k \left(\e^{\gamma_{\max}-\gamma_{\min}}-1 \right)\, .
\end{align}
The trace distance between the reduced dephased decoupled rectangular state $\omega_\sqcap^{S(0)}$
 and the Gibbs state is thus bounded from above by
\begin{equation}
  \label{eq:bound-spectrum-distance}
  \tracedistance(\omega_\sqcap^{S(0)},\rhogibbs)\le\frac{1}{2}\left(\e^{\gamma_{\max}-\gamma_{\min}}-1\right) + C\, ,
\end{equation}
where $C$ is exponentially small in the bath size.
We explicitly bound $\gamma_{\max}-\gamma_{\min}$ for a specific bath of uncoupled spins in Appendix~\Appendixdensityinfreemodel{}.

\section{\Appendixdensityinfreemodel{} Reduced density matrix of the dephased rectangular decoupled state -- a specific model}
In order to have a more explicit expression for the right hand side of Eq.~\eqref{eq:bound-spectrum-distance} let us consider a particular model for the bath.
We will also use this bath for the algorithm presented in the last part of the article.

We start with the natural choice of $m$ non-interacting uncoupled spin-$1/2$ particles with energies $0$ and $\eta$.
The spectrum of this model is discrete and the eigenvalues are integer multiples of $\eta$ between $0$ and $\norminf{\haH_B} = \eta\,m$.
The system is highly degenerate and the number of states with energy $\eta\,k$ follows a binomial distribution $\binom{m}{k}$.

This degeneracy makes it impossible to find a sufficiently good smoothed approximation $\Xi_\Delta^B(E)$ for the number of states $\Omega_\Delta^B(E)$ such that even for intervals whose width $\Delta$ scales like $d_B^{-\kappa}$ with $0<\kappa<1$ the error in the approximation of the local density $(\Xi_\Delta^B(E) - \Omega_\Delta^B(E))/d$ goes to zero for large $d_B$.
That is, the smoothed approximation would cause additional errors that would not go down exponentially fast with the bath size.

Therefore, we need to assume that the degeneracy of the levels are lifted by a suitable perturbation.
As we are only concerned about the spectrum and not the eigenstates of the bath, essentially any perturbation of adequate strength, such as basically any arbitrary weak interaction, will be sufficient.

A convenient way to perturb the model for a theoretical study is to replace the fixed local field strength by a normal distributed random field strength such that on average the energy of the local excited state is still $\eta$ above the local ground state.
If the width of the normal distribution is chosen in a suitable way, we get, with overwhelmingly high probability, a number of states $\Omega_\Delta^B(E)$ in the interval $[E,E+\Delta]$ that can be well approximated by 
\begin{equation}
  \label{eq:formulaforthedensityofstatesofourbath}
  \Xi_\Delta^B(E) = \int_E^{E+\Delta} \varrho^B(E') \dd E' \, 
\end{equation}
in the aforementioned sense, where 
\begin{equation}
  \label{eq:dos-bathfirstintroduced}
  \varrho^B(E') \coloneqq \frac{1}{\eta} 2^m\left({\frac{2}{\pi m}}\right)^{1/2} \e^{-2m(\frac{E'}{\eta m}-\frac{1}{2})^2} .
\end{equation}

Our aim is to bound $\gamma_{\max}-\gamma_{\min}$ in Eq.~\eqref{eq:bound-spectrum-distance} 
for this model. To do this, we first consider the case $\Delta \ll \norminf{\haH_B}$ in which we can approximate $\Xi_\Delta^B(E)$ as follows
\begin{equation}
  \label{eq:approxformulaforthedensityofstatesofourbath}
  \Xi_\Delta^B(E) \approx \frac{\Delta}{\eta} 2^m\left({\frac{2}{\pi m}}\right)^{1/2} \e^{-2m(\frac{E'}{\eta m}-\frac{1}{2})^2}\, .
\end{equation}
We can easily compute the second derivative of the logarithm of the above expression and find
\begin{equation}
  \frac{\partial^2 s}{\partial E^2}= -\frac{4}{\eta^2 m} \, .
\end{equation}
If $\Delta$ is not much smaller than $\norminf{\haH_B}$, $s(E)$ deviates from the parabolic form.
But, larger $\Delta$ only make the curvature of $s(E)$ smaller. In the general case we thus have
\begin{equation}
  0 > \frac{\partial^2s}{\partial E^2} > -\frac{4}{\eta^2 m}\, ,
\end{equation}
and since $\gamma_{\max}-\gamma_{\min} \leq \gamma_{\max} \leq 2\norminf{\haH_S}^2/(\eta^2 m)$ we arrive at
\begin{equation}
  \label{eq:countingbound}
  \tracedistance(\omega_\sqcap^{S(0)},\rhogibbs)\le\frac{1}{2}\left(\e^{2 \frac{\norminf{\haH_S}^2}{\eta^2 m}}-1\right)\, + C ,
\end{equation}
where $C$ is exponentially small in the bath size.
In natural situations the right hand side of the above equation will be very small as $\norminf{\haH_S}^2 \ll \eta^2 m$.

\section{\AppendixRectangularStates{} Rectangular states}
The rectangular states are a quite artificial class of states.
So what happens if instead of $\omega_\sqcap^{S(0)}$ we take states with a different energy distribution?
Will we still get something close to a Gibbs state as long as the width of the energy distribution is not too small and the density of states is exponentially increasing?
The exponential increase in the density of states ensures that most of the contribution to $\omega_\sqcap^{(0)}$ comes from the upper edge of the interval $E+\Delta$ in the case of a rectangular distribution.
Thus, it can be expected that one will get a state that is close to a Gibbs state for any energy distribution that has a region with a bounded variation followed by a sufficiently sharp cutoff to higher energies.
Where by sufficiently sharp we mean that it must drop from a value much larger than $1/d$ to a value much smaller than $1/d$ in an energy interval that is small compared to $(\partial^2 s/\partial E^2)^{-1/2}$.
The bounded variation is required as for certain systems that violate the so called \emph{eigenstate thermalization hypothesis} (ETH) \cite{NumericalQuench} details of the energy distribution can have a huge impact on certain properties of the state.
That bounded variation of the energy distribution is required to guarantee thermalization in such systems can be seen for example in the model studied in Ref.~\cite{absence}.

\section{\AppendixSystemvsBath{} Complexity of the algorithm at low temperatures}
\label{app:system-vs-bath}
In this section we discuss the average number of times the algorithm must be run before the energy window corresponding to the desired temperature is hit.

As we are only interested in the number of repetitions we can assume that the algorithm runs ``perfectly'' in that it generates exactly a rectangular state with fixed width $\Delta$ at some position $E$ of the spectrum.
Furthermore, we are only interested in cases where the reduced state of the rectangular state can be guaranteed to be close to a Gibbs state.
Thus, we may assume that the spectrum of the bath is dense enough such that the number of states of the bath in the interval $[E,E+\Delta]$ can be well approximated by a continuous and twice differentiable function $\Xi_\Delta^B(E)$ that increases exponentially to higher energies.

The probability of getting a rectangular state at energy $E$ is given by
\begin{equation}
  \label{eq:probtogetacertainE}
  P(E)=\frac{\Xi_\Delta(E)}{d} = \frac{1}{d} \sum_{k=1}^{d_S} \Xi_\Delta^B(E-E_k^{S}) \, ,
\end{equation}      
where $d=d_S d_B$ is the dimension of the Hilbert space of the total system.
Using that the density of states of the bath can be locally approximated by an exponential, we obtain
\begin{equation}
	P(E) \approx \frac{1}{d} \sum_{k=1}^{d_S} \e^{-\beta E_k^{S}}\Xi_\Delta^B(E)\, .
\end{equation}
Then, the probability of getting energy $E$ can be bounded from above by
\begin{align}
  P(E) &\approx \frac{1}{d_S} \sum_{k=1}^{d_S} \e^{-\beta E_k^{S}} \frac{\Xi_\Delta^B(E)}{d_B} \\
  &\le \frac{1}{d_S} + \e^{-\beta (E_2^S - E_1^S)} \frac{\Xi_\Delta^B(E)}{d_B} \le \frac{1}{d_S} + \e^{-\beta (E_2^S - E_1^S)}\, , \nonumber
\end{align}
with $E_2^S - E_1^S$ the gap of the system.
The number of times that the program must be run on average in order to get the rectangular state at the position $E$ is thus lower bounded by
\begin{equation}
  \label{eq:averagenumberofrunsintermsoftheprobabilitytogetacertainE}
  \overline{ \sharp \textrm{ runs} } =  \frac{1}{P(E)} \gtrapprox \frac{1}{\frac{1}{d_S} + \e^{-\beta (E_2^S - E_1^S)}} \, .
\end{equation}
This last equation shows that it is exponentially hard to go to low temperatures.
This is to be expected, as no structure of the Hamiltonian is being used and generic local Hamiltonian systems can presumably not be efficiently cooled arbitrarily close to their ground state, because this would efficiently solve {\tt QMA}-hard problems on a quantum computer \cite{localhamiltonianproblem}.
Needless to say, for specific Hamiltonians, with some additional structure, an algorithm can well be more efficient.
The coefficient of the exponential is related to the features of the spectrum at low energies. 
In the case in which there is a gap, the ground state could encode the solution of a satisfiability problem that is expected to be hard to solve.

\section{\AppendixTemperatureInTheAlgorithm{} Temperature and its error for the bath described in Appendix~\Appendixdensityinfreemodel{}}
In this section we calculate how the inverse temperature $\beta$ and its error $\delta\beta$ depend on the energy value obtained while running the algorithm and the number of qubits $q$ of register $R$ that are measured.

After running the algorithm a value $s_*$ is obtained and a state close to the rectangular state at energy $E = \norminf{H} 2^{-q} s_*$ is prepared. 
The system is then thermalized at an inverse temperature 
\begin{align}
  \beta =\frac{\dd \ln \Xi_\Delta^B(E)}{\dd E} \, .
\end{align}
Inserting the formula for $\Xi_\Delta^B(E)$ for the bath described in Appendix~\Appendixdensityinfreemodel{} \eqref{eq:formulaforthedensityofstatesofourbath} we find
\begin{align}
	\beta &=\frac{4}{\eta}\left(\frac{1}{2}-\frac{E}{\eta m}\right) = \frac{4}{\eta}\left(\frac{1}{2}-\frac{E}{\norminf{H_B}}\right)\\
	&=\frac{4}{\eta}\left(\frac{1}{2}- 2^{-q} s_* \left(1+\frac{\norminf{\haH_S}}{\norminf{\haH_B}}\right)\right)\, .
\label{eq:invT-spins}
\end{align}


The algorithm will be run repeatedly until the value $s_*$ corresponding to the desired $\beta$ is obtained. 
For the bath considered here this value is
\begin{equation}
 \label{eq:varphi}
  s_* = \left\lfloor \frac{2^q}{1+\frac{\norminf{\haH_S}}{\norminf{\haH_B}}}
  \left(\frac{1}{2} -\frac{\eta \beta}{4}\right)\right\rfloor \, .
\end{equation}
It follows from Eq.~\eqref{eq:varphi} that
\begin{equation}
  \delta\beta \leq \frac{2^{2-q}}{\eta} \left( 1 +\frac{\norminf{H_S}}{\norminf{H_B}}\right) .
\end{equation}
To reach a given precision $\delta\beta$ in the temperature it is thus sufficient to choose 
\begin{equation}
  q = \left\lceil \log_2 \frac{\norminf{\haH}}{\Delta}\right\rceil 
  =\left\lceil -\log_2\left(\frac{\delta \beta\,\eta }{1+\frac{\norminf{\haH_S}}{\norminf{\haH_B}}} \right) + 2 \right\rceil \, .
\end{equation}

\section{\Appendixalgorithconvergence{} Error and average number of runs of the quantum algorithm}
\label{app:error-algorithm}
In this section the error of the quantum algorithm is derived and a bound on the average number of repetitions that are necessary to reach the desired energy is given.
The total error is bounded by the sum of two errors 
\begin{equation}
  \tracedistance(\omega_{QC}^S,\rhogibbs) \le \tracedistance(\omega_{QC},\omega_{\sqcap}^{(0)}) + \tracedistance(\omega_{\sqcap}^{S(0)},\rhogibbs) .
\end{equation}
The deviation of the reduced rectangular state from the Gibbs state $\tracedistance(\omega_{\sqcap}^{S(0)},\rhogibbs)$ has been calculated already in Appendix~\Appendixcountingrigorous{} and \Appendixdensityinfreemodel{}
\begin{align}
  \tracedistance(\omega_\sqcap^{S(0)},\rhogibbs) &\le\frac{1}{2}\left(\e^{2 \frac{\norminf{\haH_S}^2}{\eta^2 m}}-1\right) + C .
\end{align}
Here we bound the deviation $\tracedistance(\omega_{QC},\omega_{\sqcap}^{(0)})$ of the final state of the algorithm from the rectangular state.
Our final result for the bath discussed in Appendix~\Appendixdensityinfreemodel{} will be
\begin{align}
  \tracedistance(\omega_{QC},\omega_{\sqcap}) &\le \e^{\frac{2\norminf{\haH_S}^2}{\eta^2 m}+\beta \norminf{\haH_S}+\frac{\eta^2 m \beta^2}{8} } \\
  & \times 2^{q-r+2} \left(1+\ln(2^{r-q})/\pi^2 \right) + C\nonumber 
\end{align}
(again $C$ is exponentially small in the bath size).
It is convenient to write $\eta$ in terms of a dimensionless parameter $\lambda > 0$
\begin{equation}
  \eta=\left({\frac{\lambda}{m}}\right)^{1/2} \norminf{\haH_S} \, ,
\end{equation}
because then the two errors can be written in the form:
\begin{align}
  \tracedistance(\omega_\sqcap^{S(0)},\rhogibbs) &\le\frac{1}{2}(\e^{\frac{2}{\lambda}}-1) + C \\
  \tracedistance(\omega_{QC},\omega_{\sqcap}) &\le \e^{\frac{2}{\lambda}+\beta \norminf{\haH_S}+\frac{\lambda \norminf{\haH_S}^2 \beta^2}{8} } \\
  & \times 2^{q-r+2} \left(1+\ln(2^{r-q})/\pi^2 \right) + C \nonumber
\end{align}

To reduce the trace distance $\tracedistance(\omega_\sqcap^{S(0)},\rhogibbs)$ it is favorable to choose a large $\lambda$.
Intuitively this captures the fact that the energy content in the bath must be much larger than that of the system in order to get a Gibbs state.
However, at the same time this increases the error in the approximation of the rectangular state, because 
performing the phase estimation on a system with a more dense spectrum is harder.
This increase of the error can however be compensated by increasing the accuracy of the phase estimation by linearly increasing the number of qubits $r$ in the ancilla register.
Similarly, lower temperatures can be reached by a quadratic increase of the number of ancilla qubits of order $O(\beta^2\norminf{\haH_S}^2)$.
As the phase estimation requires a total number of $2^r$ controlled unitary operations, this comes at an exponential cost in terms of runtime.
This is expected to be a problem common to all algorithms using phase estimation on many particle systems.

Before we go into the derivation of $\tracedistance(\omega_{QC},\omega_{\sqcap})$, let us quickly give the average number of runs that are required to get a certain temperature for the bath discussed in Appendix~\Appendixdensityinfreemodel{} in terms of $\lambda$.
From Eqs.~\eqref{eq:probtogetacertainE} and \eqref{eq:averagenumberofrunsintermsoftheprobabilitytogetacertainE} and the explicit formula for the smoothed number of states given in Eq.~\eqref{eq:explicitesmoothednumberofstates} it is easy to see that
\begin{equation}
  \overline{ \sharp \textrm{ runs} } \leq \sqrt{\frac{\pi \lambda}{2}} \frac{\norminf{\haH_S}}{\Delta} \e^{\frac{2}{\lambda}+\beta \norminf{\haH_S}+\frac{\lambda \norminf{\haH_S}^2 \beta^2}{8} } .
\end{equation}

\subsection{Deviation from the rectangular state for a general bath}

Here we bound the trace distance $\tracedistance(\omega_{QC},\omega_{\sqcap})$ between the state $\omega_{QC}$ generated by the circuit with $r$ ancilla qubits when, after measuring $q$ of them, the outcome $s_*$ is obtained, and the rectangular state $\omega_{\sqcap}$ in the interval $[E,E+\Delta]$ with $E = s_* 2^{-q} \norminf{\haH}$ and $\Delta=2^{q-r} \norminf{\haH}$ is prepared.

\begin{figure}[tb]
  \includegraphics[width=0.8\linewidth]{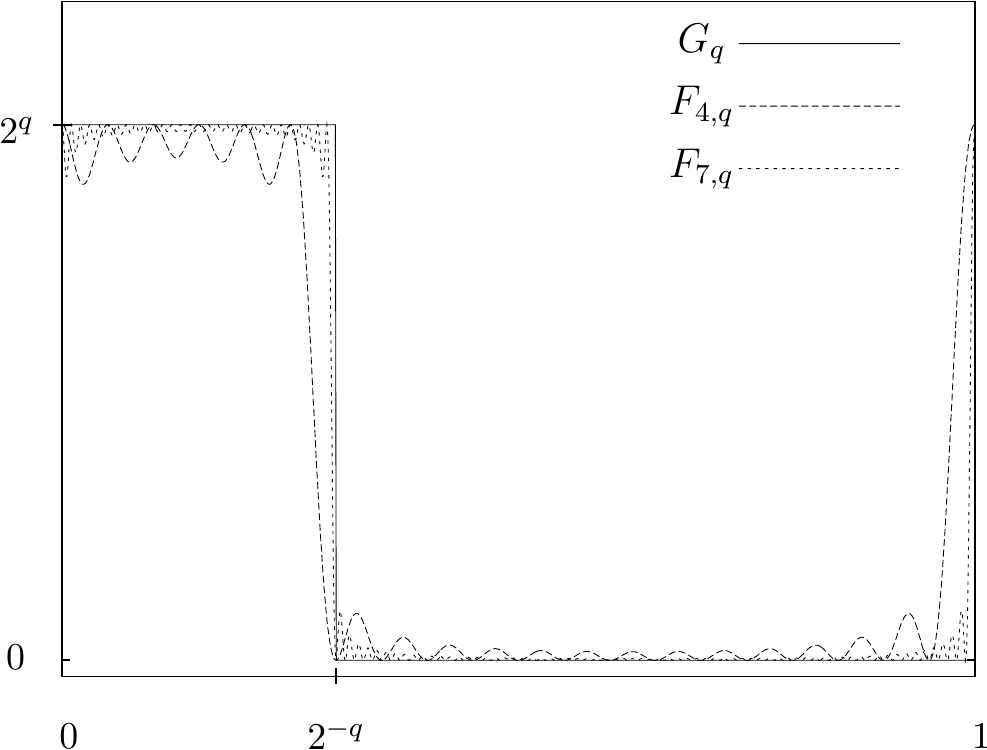}
  \caption{\label{fig:qpe-circuit}Plot of the $G_q$ and $F_{r,q}$ functions. As $r$ increases, $F_{r,q}$ tends to the rectangular function $G_q$. }
\end{figure}

This trace distance can be written in terms of the one norm distance of the distributions of the diagonal elements
\begin{equation}
\label{eq:error-circuit}
    \tracedistance(\omega_{QC},\omega_{\sqcap}) = 
    \frac{1}{2} \sum_{k=1}^d \left| q_k - p_k \right|\, ,
\end{equation}
where the $\{q_k\}$ and $\{p_k\}$ are the eigenvalues of $\omega_{QC}$ and $\omega_{\sqcap}$ respectively, which can be written as
\begin{align}
q_k&=\frac{F_{r,q}(\varphi_k-\tilde \varphi)}{Z_F}, \\
p_k&=\frac{G_q(\varphi_k-\tilde \varphi)}{Z_G}\, ,
\end{align}
where $\varphi_k \coloneqq  E_k/\norminf{\haH}$ and $\tilde\varphi \coloneqq E/\norminf{\haH}$ are the phases corresponding to the eigenvalues of the Hamiltonian and the energy of the rectangular state,
$Z_F\coloneqq \sum_{j=1}^{d}F_{r,q}(\varphi_j-\tilde \varphi)$ and $Z_G\coloneqq \sum_{j=1}^d G_q(\varphi_j-\tilde \varphi)$ are normalization constants, and the functions $F_{r,q}$ and $G_q$ are defined as
\begin{align}
  \label{eq:defofFN}
  F_{r,q}(\varphi)&\coloneqq 2^{q-r} \sum_{k=0}^{2^{r-q}} f_{r}\left( \varphi-k\,2^{-r} \right) ,\\
  \label{eq:defofG}
  G_q(\varphi)&\coloneqq 2^q \left(\Theta (\varphi) -\Theta (\varphi-2^{-q})\right) \, ,
\end{align}
with
\begin{eqnarray}
	f_r(\varphi)=2^r |\alpha_0(\varphi)|^2 = 2^{-r} \frac{\sin^2(\pi 2^r \varphi)}{\sin^2 (\pi \varphi)}, 
\end{eqnarray}	
and $\Theta$ the step function (see main text and Fig.~\ref{fig:qpe-circuit}).

Eq.~\eqref{eq:error-circuit} is bounded by 
\begin{equation}
\tracedistance(\omega_{QC},\omega_{\sqcap}) 
\le  \frac{1}{Z_G} \sum_{k=1}^d \left| F_{r,q}(\varphi_k-\tilde \varphi) - G_q(\varphi_k-\tilde \varphi) \right|\, ,
\label{eq:sumF-G}
\end{equation}
where we have used that
\be
\left\vert \frac{F_{r,q}(x)}{Z_F}-\frac{G_{q}(x)}{Z_G}\right\vert \le  \frac{\vert F_{r,q}(x)-G_q(x)\vert}{Z_G}
+F_{r,q}(x)\left\vert \frac{1}{Z_F} -\frac{1}{Z_G}\right\vert
\nonumber
\ee
and
\be
\vert Z_F-Z_G\vert \le \sum_{k=1}^d \left| F_{r,q}(\varphi_k-\tilde \varphi) - G_q(\varphi_k-\tilde \varphi) \right| \, .
\nonumber
\ee
Notice that $Z_G= \Omega_\Delta(E) 2^q $, where $\Omega_\Delta(E)$ is the number of states of the total system in the interval $[E, E+\Delta]$.

Using that the spectrum of the whole system is sufficiently dense it is possible to approximate the upper bound in Eq.~\eqref{eq:sumF-G} by an integral.
To make this rigorous we decompose the sum \eqref{eq:sumF-G} into bins of width $\norminf{\haH} / L$,
\begin{equation}
	\tracedistance(\omega_{QC},\omega_{\sqcap}) \le
	\sum_{j=0}^{L-1} \sum_{E_k\in \left[\frac{j}{L},\frac{j+1}{L}\right[}h(E_k)\, ,
\label{eq:distance-H}
\end{equation}
where
\begin{equation}
  \label{eq:h-definition}
  h(E') \coloneqq \frac{1}{Z_G}
  \left| F_{r,q}\left(\frac{E'}{\norminf{\haH}}-\tilde\varphi \right) -G_q\left(\frac{E'}{\norminf{\haH}}-\tilde\varphi\right)  \right|  \, 
\end{equation}
is a function introduced to simplify the notation and $L$ is the number of bins in
which the spectrum has been divided.
The idea is to take an $L$ as large as possible such that the number of energy values in a bin $\Omega_{\norminf{\haH}/L} (j \norminf{\haH}/L)$ can still be well approximated by its smoothed version $\Xi_{\norminf{\haH}/L} (j \norminf{\haH}/L)$ (compare the discussion of the twice differentiable approximation in Appendix~\Appendixcountingrigorous{} and \Appendixdensityinfreemodel{}).
For baths for which such a smooth approximation is possible, one can always take an $L$ proportional to the dimension of the Hilbert space $d$ or, at least, to some power $d^\kappa$ with $0 < \kappa < 1$ such that $L$ scales exponentially with the size of the bath.

As $h$ is continuos differentiable except at $E$ and $E+\Delta$, Taylor's theorem ensures that for all bins $j$ (except for the two bins that contain $E$ and $E+\Delta$, which we can simply ignore because they only contribute an error of order $1/L$) there exists some value $\xi_j \in [j \norminf{\haH}/L, (j+1)\norminf{\haH}/L[$ for which
\begin{equation}
  h(E_k) = h\left( \frac{j \norminf{\haH}}{L}\right)+ h'( \xi_j )\left(E_k-\frac{j \norminf{\haH}}{L}\right)\, ,
\label{eq:H-dif}
\end{equation}
where $j \norminf{\haH}/L\le E_k < (j+1)\norminf{\haH}/L$.
Then, the contribution of the $j$-th bin in Eq.~\eqref{eq:distance-H} can be bounded from above by
\begin{align}
\label{eq:bound-bin}
 \sum_{E_k\in \left[\frac{j}{L},\frac{j+1}{L}\right[}&h(E_k) \le \Omega_{\frac{\norminf{\haH}}{L}} 
  \left(\frac{j \norminf{\haH}}{L}\right) \\
	&\times \left( h\left(\frac{j \norminf{\haH}}{L}\right)+
	\frac{1}{L} \sup_{\xi_j\in\left[\frac{j}{L},\frac{j+1}{L}\right[}h'(\xi_j) \norminf{\haH}\right)\nonumber \, .
\end{align}
The last term in the parenthesis in Eq.~\eqref{eq:bound-bin} decreases with $L$ and is thus exponentially small in the bath size.
This can be verified with a lengthy, but straightforward calculation.

Notice that the number of states can be written as
\begin{equation}
  \Omega_{\frac{\norminf{\haH}}{L}} \left(E'\right) = \varrho \left(E'\right) \frac{\norminf{\haH}}{L} + O(L^{-2})\, .
\label{eq:Omega-vs-rho}
\end{equation}
where 
\begin{equation}
  \varrho =\frac{\dd \Xi_\Delta(E')}{\dd E'}\, .
\end{equation}

Putting Eqs.~\eqref{eq:distance-H}, \eqref{eq:bound-bin} and \eqref{eq:Omega-vs-rho} together,
the error of the circuit can be bounded from above by
\begin{align}
	\tracedistance(\omega_{QC},\omega_{\sqcap}) &\le \sum_{i=0}^{L-1} h\left(\frac{i \norminf{\haH}}{L}\right)
	\varrho\left( \frac{i \norminf{\haH}}{L} \right)\frac{\norminf{\haH}}{L}\nonumber \\
	&+O\left(L^{-1}\right)\, .
\end{align}
Notice that the upper bound of the previous equation converges to an integral for $L\to \infty$ with deviations
that scale as $O(L^{-1})$.
Thus, the trace distance between $\omega_{QC}$ and $\omega_\sqcap$ is bounded from above by
\begin{equation}
	\tracedistance(\omega_{QC},\omega_{\sqcap})\le \normone{h \, \varrho} +O\left(L^{-1}\right)\, ,
	\label{eq:td-circuit-1}
\end{equation}
where 
\begin{equation}
  \normone{h \, \varrho}=
  \int_0^{\norminf{\haH}}  h(E') \varrho(E') \dd E' \, .
\end{equation}	
We will bound $\normone{h \, \varrho}$ for the model described in Appendix \Appendixdensityinfreemodel{} in the next section.

\subsection{Deviation from the rectangular state for the bath described in Appendix \Appendixdensityinfreemodel{}}
Next, the error of the circuit for a bath described in Appendix \Appendixdensityinfreemodel{} is bounded.
The one norm $\normone{h\, \varrho}$ from Eq.~\eqref{eq:td-circuit-1} can be bounded by
\begin{equation}
  \normone{h\, \varrho} \le   \frac{2^{-q}}{\Xi_\Delta(E)}\normone{F_{r,q}-G_q}\sup_{0\le E' \le \norminf{\haH}}\varrho(E') \, ,
\end{equation}
where the definition of $h$ given in Eq.~\eqref{eq:h-definition} has been used.
We now find an upper bound on the supremum and a lower bound on the smoothed number of states.

The density of states of the total system can be written in terms of the density of states of the bath as
\begin{equation}
  \varrho(E')=\sum_{k=1}^{d_S} \varrho^B(E'-E_k^S) .
  \label{eq:total-dos}
\end{equation}
Inserting the density of states of our bath \eqref{eq:dos-bathfirstintroduced} in the previous
expression and writing it as a function of the inverse temperature given in Eq.~\eqref{eq:invT-spins} we get
\begin{align}
  \varrho(E)&= \sum_{k=1}^{d_S}  \e^{-\frac{2(E_k^S)^2}{\eta^2 m}}
  \e^{-\beta E_k^S}  \nonumber \\
  &\times d_B \frac{1}{\eta} \left(\frac{2}{\pi m}\right)^{1/2}  \e^{-\eta^2 m \beta^2/8 }\, .
\end{align}
As we only consider positive temperatures $\beta\geq0$ it is easy to see that the density of states attains its maximum value at $\beta=0$, thus
\begin{equation}
  \sup_{0\le E' \le \norminf{\haH} } \varrho(E') \leq \frac{d}{\eta} \left({\frac{2}{\pi m}}\right)^{1/2} \, ,
\end{equation}
where $d=d_S\,d_B$ is the dimension of the Hilbert space of the total system.

The smoothed number of states $\Xi_\Delta(E)$ can be lower bounded as follows
\begin{align}
  \label{eq:explicitesmoothednumberofstates}
  \Xi_\Delta(E)&=\int_E^{E+\Delta}\varrho(E') \dd E'\ge \varrho(E)\Delta  \\
  &\ge \frac{\Delta  d}{\eta} \left({\frac{2}{\pi m}}\right)^{1/2}
  \e^{\frac{-2\norminf{\haH_S}^2}{\eta^2 m}-\beta \norminf{\haH_S}-\frac{\eta^2 m \beta^2}{8} }\, ,\nonumber
\end{align}
where again we have assumed that $E$ is in a position of the spectrum with positive temperature.

Finally we obtain
\begin{align}
  \label{eq:error-rectangulars}
  \tracedistance(\omega_{QC},\omega_{\sqcap})&\le  
  \e^{\frac{2\norminf{\haH_S}^2}{\eta^2 m}+\beta \norminf{\haH_S}+\frac{\eta^2 m \beta^2}{8} }
  \frac{\normone{F_{r,q}-G_q}}{\norminf{\haH}}\nonumber \\
  & + O(L^{-1})\, .
\end{align}
In order for this error to become small the one norm $\normone{F_{r,q}-G_q}$ must be made small enough such that it compensates the exponential prefactors.
As we will see in the next section this can be achieved by using a polynomially large ancilla register $R$.

\subsection{One norm between $F_{r,q}$ and $G_q$}
In this section we bound the one norm between $F_{r,q}$ and $G_q$ introduced in Eqs.~\eqref{eq:defofFN} and \eqref{eq:defofG}.
The one norm between $F_N$ and $G$ is defined as 
\begin{equation}
  \normone{F_{r,q}-G_q}=\int_0^{\norminf{\haH}}\left| F_{r,q}\left(\frac{E'}{\norminf{\haH}}\right)-
    G_q\left(\frac{E'}{\norminf{\haH}}\right) \right| \dd E' . \nonumber
\end{equation}
By a simple change of variables it is easy to show that
\begin{equation}
	\frac{\normone{F_{r,q}-G_q}}{\norminf{\haH}}=\int_0^1\left| F_{r,q}(\varphi)-G_q(\varphi) \right| \dd \varphi \, ,
\end{equation}
Remember that both $F_{r,q}$ and $G_q$ are normalized on the interval $[0,1]$ and that $G_q(\varphi)$ is a step function that is non-zero only in the interval $[0,2^{-q}[$, and that in this interval $F_{r,q}(\varphi)<G_q(\varphi)=2^q$.
Using this, the previous integral can be rewritten as
\begin{align}
	\frac{\normone{F_{r,q}-G_q}}{\norminf{\haH}}&=2\int_{2^{-q}}^1 F_N(\varphi) \dd \varphi \\
	&=2^{q-r+1} \sum_{k=0}^{2^{r-q}}\int_{2^{-q}}^1 f_r\left( \varphi - k\, 2^{-r} \right) \dd \varphi \, .\nonumber
\end{align}
Due to the symmetry and periodicity of $f_r$, the contribution to the previous integral of the right tail 
of $f_r(\varphi-k\,2^{-r})$ is the same as the contribution of the left tail of $f_r(\varphi-(2^{-q} - k\,2^{-r}))$ for $k\,2^{-r} \le 2^{-q}$, thus
\begin{equation}
  \begin{split}
    &\int_{2^{-q}}^{k\,2^{-r}+\frac{1}{2}} f_r\left( \varphi-k\,2^{-r}\right) \dd \varphi\\
    = &\int_{2^{-q}-k\,2^{-r}+\frac{1}{2}}^1 f_r\left( \varphi-2^{-q}+k\,2^{-r}\right) \dd \varphi . 
  \end{split}
\end{equation}
This implies that
\begin{equation}
	\frac{\normone{F_{r,q}-G_q}}{\norminf{\haH}}
	=2^{q-r+2} \sum_{k=0}^{2^{r-q}}\int_{2^{-q}}^{k\,2^{-r}+\frac{1}{2}} f_r\left( \varphi-k\,2^{-r}\right) \dd \varphi \, .
\end{equation}
The integral of the previous equation can be bounded by
\begin{equation}
\int_{2^{-q}}^{k\,2^{-r}+\frac{1}{2}} f_r\left( \varphi-k\,2^{-r}\right) \dd \varphi \le \frac{\cot \left(\pi (2^{-q}-k\,2^{-r})\right)}{\pi\,2^r} \, ,
\end{equation}
where the bound 
\begin{equation}
  f_r\left( \varphi-k\,2^{-r}\right) \le \frac{2^{-r}}{\sin^2\left( \varphi-k\,2^{-r}\right)} 
\end{equation}
has been used.
Therefore, the distance between the functions $F_{r,q}$ and $G_q$ can be bounded by
\begin{equation}
\frac{\normone{F_{r,q}-G_q}}{\norminf{\haH}}
\le 2^{q-r+2} \left(\frac{1}{2} + \sum_{k=0}^{2^{r-q}-1} \frac{\cot \left(\pi (2^{-q}-k\,2^{-r})\right)}{\pi 2^{r}} \right)\, ,
\end{equation}
where the sum has been split up in the $k=2^{r-q}$ case, which is exactly $1/2$, and the rest. 
The sum of the previous equation can be bounded in two steps by 
\begin{align}
\sum_{k=1}^{2^{r-q}} &\pi\, 2^{-r} \cot( \pi k 2^{-r} ) \nonumber \\
&\le \pi\,2^{-r} \cot(\pi\,2^{-r}) + \int_\frac{1}{N}^{c} \cot (\pi u) \dd u \\
&\le 1+ \ln(2^{r-q}) \, .
\end{align}
Thus, the one norm between $F_N$ and $G$ is bounded by
\begin{equation}
\frac{\normone{F_N-G}}{\norminf{\haH}}\le 2^{q-r+2} \left(\frac{1}{2}+\frac{1}{\pi^2}+\frac{\ln(2^{r-q})}{\pi} \right)\, .
\end{equation}
Inserting this into Eq.~\eqref{eq:error-rectangulars}, we finally get
\begin{align}
  \tracedistance(\omega_{QC},\omega_{\sqcap}) & \le \e^{\frac{2\norminf{\haH_S}^2}{\eta^2 m}+\beta \norminf{\haH_S}+\frac{\eta^2 m \beta^2}{8} } \\
  & \times 2^{q-r+2} \left(1+\ln(2^{r-q})/\pi^2 \right) + C \, , \nonumber
\end{align}
where $C$ is exponentially small in the bath size.
This completes the discussion of an upper bound of the error made in the
quantum algorithm.

\section{\Appendixboixo{} Discussion of the argument presented in Ref.~\cite{bilgin}}
\label{app:boixo}
Ref.~\cite{bilgin} presents a novel approach towards thermalizing quantum systems using an iterative approach,
in which pre-thermalized parts are put together in a suitable fashion in order to arrive at a Gibbs state of a 
quantum system with a local Hamiltonian. This argument provides a new intuition on how one can think
of thermalizing local quantum systems, different from a quantum 
Monte Carlo approach (and the one presented here).
In this appendix, however, we point out a subtle mistake of the published version of the paper; however,
this challenge can be overcome and a new proof can be formulated that retains key elements of the 
previous structure of the argument \cite{Private}. The intuition behind the argument and in fact the algorithm
itself hence turn out to be correct.

Each merging step consists of two steps. The first is a probabilistic step that updates the probability
weights of the Gibbs state by means of postselection. The second one aims at rotating 
the eigenbasis of the old Hamiltonian to the one of the new Hamiltonian by means of an instance of dephasing.
Each step has as input a chosen $\epsilon>0$.
For the entire algorithm to work, this procedure has to be correct up to errors of $O(\epsilon^2)$.
In what follows we refer to the equation numbering of the preprint v2.

In Eq.~(5), perfect dephasing is being achieved when $\sigma\rightarrow\infty$. This is
approximated by imperfect dephasing with a finite $\sigma$. From the Dyson series
of second order (6)-(8), it follows that (in the asymptotic notation) 
\begin{equation}
	\sigma = O(1/\epsilon).
\end{equation}
This is not made explicit in the 
paper, but appears to be crucial.
Intuitively speaking, if $\sigma$ becomes larger, the dephasing is more exact, but then (6)-(8) can 
no longer be used.

This step is then used in the procedure following Eq.~(13). A $\zeta$ is introduced and dephasing
between eigenstates with relative gap larger than $\zeta$ is considered. Then a new
Hamiltonian $\tilde{\haH}$ is constructed, which has the following feature: It has the same eigenbasis as $\haH+\epsilon h$, but with eigenvalues grouped in bins, such that the smallest gap between bins is $\zeta$. 
For the following procedure to work, one has to take (again in the asymptotic notation)
\begin{equation}
	\sigma = \theta(1/\zeta), 
\end{equation}
so both $\sigma = O(1/\zeta)$ and $\sigma = \Omega(1/\zeta)$ 
(although only $\sigma = O(1/\zeta)$ is made explicit, both bounds are actually needed).

This, however, seems to already fix $\zeta = \Omega(\epsilon)$. So there is no longer the 
freedom to have
\begin{equation}
	\zeta= \epsilon^2 \beta \|h\|^2,
\end{equation}
which would mean that $\zeta= \theta(\epsilon^2)$. This appears to contradict the 
above statement. Again, intuitively,  \emph{$\sigma$ is forced to be small for the Dyson approach to work},
but at the same time this gives a constraint to $\zeta$ which \emph{requires $\sigma$ to be large}.

\section{\Appendixtheoremonefornonconstantdensityofstates{} Theorem~1 for an exponential density of states}
As we have seen in the main text in theorem~1 
the distinguishability of the microcanonical states $\omega_\sqcap^{(0)}$ and $\omega_\sqcap$ corresponding to an interval $[E,E+\Delta]$ of the Hamiltonians $\haH_0$ and $\haH = \haH_0 + V$ is bounded by
  \begin{align}
\label{eq:theorem1-in-appendix}
    \tracedistance(\omega_\sqcap^{S},\omega_\sqcap^{S(0)}) \leq \tracedistance(\omega_\sqcap,\omega_\sqcap^{(0)}) \leq \frac{\norminf{V}}{\varepsilon} + \frac{\Delta\Omega+\Omega_\varepsilon}{2\,\Omega_{\mathrm{max}}} \, ,
  \end{align}
  where $\Omega_{\mathrm{max}}$ and $\Delta\Omega$ are the maximum, and the difference, of the dimensions of the support of $\omega_\sqcap^{(0)}$ and $\omega_\sqcap$, and $\Omega_\varepsilon$ is the total number of eigenstates of $\haH$ and $\haH_0$ in the intervals $[E,E+\varepsilon]$ and $[E+\Delta-\varepsilon,E+\Delta]$.
  
In the main text, in order to give a more comprehensible interpretation to Eq.~\eqref{eq:theorem1-in-appendix}, an approximately constant density of states was assumed and it was shown that
  \begin{equation}
    \label{eq:firstintuitiveinterpretationoftheoremone}
    \tracedistance(\omega_\sqcap^{S},\omega_\sqcap^{S(0)}) \lessapprox \frac{3\sqrt{2}}{2} \left({\frac{\norminf{V}}{\Delta}}\right)^{1/2} .
  \end{equation}  
  However, we have seen that thermal states emerge in situations where the density of states is locally well approximable  by an exponential.
Thus, new conditions that ensure the indistinguishability of the microcanonical states $\omega_\sqcap^{S}$ and 
$\omega_\sqcap^{S(0)}$ must be derived for the exponential density of states
  \begin{equation}
    \varrho(E) \propto e^{\beta E} \, .
  \end{equation}
  In order to do this, let us notice that both terms in the upper bound of 
  Eq.~\eqref{eq:theorem1-in-appendix} are positive and must be
 simultaneously and independently small. For the first term, this is only possible if $\norminf{V} \ll \varepsilon$.
To find the condition for the second term, let us assume that 
the interaction does not shift excessively the energy levels such that $\Delta \Omega / \Omega_{\mathrm{max}}$ can
be neglected. Then, 
\begin{equation}
  1\gg \frac{\Omega_\varepsilon}{2\,\Omega_{\mathrm{max}}} > 
  \frac{\int_{E+\Delta-\varepsilon}^{E+\Delta} \varrho(E')\dd E'}{2\int_{E}^{E+\Delta} \varrho(E')\dd E'}
  =  \frac{1-\e^{-\beta \varepsilon}}{2(1-\e^{-\beta \Delta})}\, ,
\end{equation}
and the condition $\beta \varepsilon \ll 1$ is required. These two necessary conditions can be summarized
 as
\begin{equation}
  \label{eq:conditions-for-v-beta}
  \beta \norminf{V} \ll \beta \varepsilon \ll 1 \, .
\end{equation}
It is easy to see that they are also essentially sufficient conditions by using
\begin{equation}
  \frac{\Omega_\varepsilon}{2\,\Omega_{\mathrm{max}}}<
  \frac{\int_{E+\Delta-\varepsilon}^{E+\Delta} \varrho(E')\dd E'}{\int_{E}^{E+\Delta} \varrho(E')\dd E'}
  =\frac{1-\e^{-\beta \varepsilon}}{1-\e^{-\beta \Delta}}
  \le\frac{\beta \varepsilon}{1-\e^{-\beta \Delta}}\, ,\nonumber
\end{equation}
to bound Eq.~\eqref{eq:theorem1-in-appendix} as follows
\begin{align}
  \tracedistance(\omega_\sqcap^{S},\omega_\sqcap^{S(0)})
  &\le \frac{\norminf{V}}{\varepsilon} + \frac{\beta \varepsilon}{1-\e^{-\beta \Delta}} + \frac{\Delta\Omega}{\Omega_{\mathrm{max}}} \\
  &\approx \frac{\norminf{V}}{\varepsilon} + \frac{\beta \varepsilon}{1-\e^{-\beta \Delta}} \, .
\end{align}

Finally, let us simply choose $\varepsilon=\sqrt{\norminf{V}/\beta}$. The trace distance between the microcanonical states $\omega_\sqcap^{(0)}$ and $\omega_\sqcap$ then reads
\begin{equation}
  \label{eq:distinguishability-with-temperature}
  \tracedistance(\omega_\sqcap^{S},\omega_\sqcap^{S(0)})\lessapprox
  \frac{2}{1-\e^{-\beta \Delta}}\sqrt{\beta\norminf{V}}\, .
\end{equation}

Equation~\eqref{eq:distinguishability-with-temperature} ensures the indistinguishability of the
interacting and non-interacting microcanonical states as long as $\Delta > k_B T$ is not too small and the condition
\be
\norminf{V}\ll k_B T \, ,
\ee
where $k_B$ is the Boltzmann constant and $T$ the absolute temperature, is fulfilled. That is,
Eq.~\eqref{eq:distinguishability-with-temperature} gives us a physical intuition about when an interaction is \emph{weak} in the sense of theorem~1:
An interaction is weak if it is small compared to the thermal energy $k_B T$, which is a measure of the
intensive energy content of the system.
In conclusion we see that how strong an interaction feels for a system depends on how much energy it contains.


\begin{thebibliography}{34}
\bibitem{footnote1}
        Reference \cite{Goldstein06} only deals with the non interacting case.
        In Ref.~\cite{tasaki98} weak means that $\norminf{V}$ is much smaller than 
        the level spacing of $\haH_S$, which is well defined in the limit of large baths, 
        but still breaks down quickly even for relatively small subsystems and 
        moreover the rigorous results are limited to cases where $V$ only couples neighboring eigenstates.
        Ref.~\cite{dong07} contains a non rigorous discussion of the general 
        case and numerical results suggesting that couplings much larger 
        than the level spacing can be sufficiently weak for Gibbs states to emerge; a fact made 
        rigorous in this work.
\end{thebibliography}
\end{document}